\documentclass{article}[12pt]
\usepackage{amsmath,amssymb}
\usepackage{graphicx}%
\usepackage{verbatim}

\newcommand{\TwoFig}[4]{%
\begin{center}
\begin{tabular}{lr}
\parbox{8cm}{\includegraphics[width=8cm]{#1}}  & \parbox{8cm}{\includegraphics[width=8cm]{#2}} \\
\parbox{8cm}{\vspace{7pt}\refstepcounter{figure}Figure \thefigure.\quad #3\vfill} & \parbox{8cm}{\vspace{7pt}\refstepcounter{figure}Figure \thefigure.\quad #4\vfill} \\
\end{tabular}
\end{center}
\vspace{12pt}
}
\newcommand{\TwoFigName}[3]{%
\begin{center}
\begin{tabular}{l}
\parbox{16cm}{\parbox{8cm}{\includegraphics[width=6cm]{#1}}  \parbox{8cm}{\includegraphics[width=6cm]{#2}}} \\
\parbox{16cm}{\vspace{7pt}\refstepcounter{figure}Figure \thefigure.\quad #3\vfill} \\
\end{tabular}
\end{center}
\vspace{7pt}
}
\newcommand{\ThreeFigName}[4]{%
\begin{center}
\begin{tabular}{l}
\parbox{16.3cm}{
\parbox{5.1cm}{\includegraphics[width=5.1cm]{#1}}  \parbox{5.1cm}{\includegraphics[width=5.1cm]{#2}}  \parbox{5.1cm}{\includegraphics[width=5.1cm]{#3}}}\\
\parbox{16.3cm}{\vspace{7pt}\refstepcounter{figure}Figure \thefigure.\quad #4\vfill}
\end{tabular}
\end{center}
\vspace{7pt}
}

\begin{document}

\begin{center}
{\bf \Large Complete model of cosmological evolution of a classical scalar field with the Higgs potential. I. Analysis of the model\\[12pt]
Yu. G. Ignat'ev, A. R. Samigullina }\\
Physics Institute of Kazan Federal University,\\
Kremleovskaya str., 35, Kazan, 420008, Russia.
\end{center}

\begin{abstract}

A complete model of cosmological evolution of a classical scalar field with the Higgs potential is studied and simulated on a computer without assumption that the Hubble constant is nonnegative. The corresponding dynamical system is qualitatively analyzed and the Einstein-Higgs hypersurfaces the topology of which determines the global properties of the phase trajectories of the cosmological model are classified.\\

\noindent
{\bf keywords}: cosmological models, Higgs field, Einstein-Higgs hypersurface, global behavior.

\end{abstract}

\section*{Introduction}
 In a number of recent works \cite{IgnSTFI}-\cite{Ign2}, one of the authors formulated the assumption on the possible existence of the so-called limiting Euclidean cycles in cosmological models based on classical and phantom scalar fields. The essence of the phenomenon of the Euclidean cycles consists in aspiration of the cosmological models with definite field model parameters to the state with zero effective energy. In this case, the space-time becomes pseudo-Euclidean, though the scalar fields are nonzero and perform free nonlinear oscillations. This assumption was based on an analysis of a great number of both qualitative and numerical studies of cosmological models with scalar fields (for example, see \cite{Ign3}-\cite{Ign4} and the references therein). Within the framework of this paper, we will not give a review of works on cosmological models with scalar fields of other authors that do not directly related to the subject of our research. Such sufficiently detailed review can be found in \cite{IgnSTFI},\cite{Ign2}. In our works \cite{IgnSam2},\cite{IgnSam3}, the phenomenon of the Euclidean cycles, in particular, the nonlinear oscillation spectrum with characteristic features of the thermal spectrum have been studied in ample details.

Note that all works mentioned above were based on dynamic systems in which the Hubble constant H(t) was substituted from Einstein's equation  $^4_4$ as its non-negative root that allowed the dimension of the dynamic system to be decreased by unity. However, in \cite{Ign5} it was shown that the\emph{ assumption about non-negative} H(t) contradicts the complete system of dynamic equations formulated in this work. The complete dynamic system was preliminary analyzed in \cite{Ign5}, in particular, its qualitative analysis was performed, and it was shown that the dynamic system based on the classical Higgs field can proceed from the expansion stage to the compression stage in the process of cosmological evolution. This absolutely unexpected property of the examined cosmological models requires additional, more detailed research. Besides, it is necessary to investigate more carefully the phase of model transition from the expansion stage to the compression stage when zero and close to zero values of the Hubble constant are realized and the geometry becomes close to pseudo-Euclidean one. The present work is devoted exactly to the study of these questions for the cosmological models based on the classical scalar Higgs field.

\section{Mathematical model}
\subsection{Basic relationships}
Let us describe, according to \cite{Ign5}, basic relationships of the mathematical model describing the cosmological evolutions of the classical scalar Higgs field and its main properties. As a field model, we consider the self-consistent system of Einstein's equations and the classical scalar field $\Phi$ with the Higgs potential characterized by the Lagrange function
\begin{equation} \label{Eq__1}
L_s=\frac{1}{8\pi}\bigg(\frac{1}{2}g^{ik} \Phi _{,i} \Phi _{,k} -V(\Phi )\bigg),
\end{equation}
where $V(\Phi)$ is the potential energy of the scalar field:
\begin{equation} \label{Eq__2}
V(\Phi )=\displaystyle -\frac{\alpha }{4} \left(\Phi ^{2} -\frac{em^{2} }{\alpha } \right)^{2}
\end{equation}
$\alpha$  is the self-action constant, \emph{m} is the scalar boson mass, and $e=\pm 1$  is the indicator. The scalar field equation
\begin{equation} \label{Eq__3}
\Delta\Phi+V'_\Phi=0
\end{equation}
where $\Delta=g^{ik}\nabla_i\nabla_k$ and the energy-momentum tensor of the scalar field
\begin{equation} \label{Eq__4}
T^i_k=\frac{1}{8\pi}\bigg(\Phi^{,i}\Phi _{,k}-\frac{1}{2}\delta^i_k\Phi _{,j}\Phi^{,j}+\delta^{i}_kV(\Phi)\bigg)
\end{equation}
correspond to the Lagrange function given by \eqref{Eq__1}. Below we omit the constant term in the Higgs potential defined by \eqref{Eq__2} that leads to a simple redefinition of the cosmological constant $\lambda$  bearing in mind that the cosmological constant $\lambda$  is renormalized as follows:
\begin{equation} \label{Eq__5}
\lambda=\lambda_0-\frac{m^4}{4\alpha}
\end{equation}
where $\lambda_0$  is its unperturbed initial value. The corresponding Einstein equations for the system under study have the form \footnote{Everywhere in the text, $G=\hbar=c=1$, the metric signature is $(-,-,-,+)$ , and the Ricci tensor is obtained through  convolution over the first and third indices}
\begin{equation} \label{Eq__6}
G_k^i\equiv R^i_k-\frac{1}{2}R\delta^i_k=8\pi T^i_k+\lambda\delta^i_k.
\end{equation}
Equations \eqref{Eq__3} and \eqref{Eq__6} are the basic equations of the model.
In \cite{Ign5} it was shown that for the spatially flat Friedman Universe
$$ds^2_0=dt^2-a^2(t)(dx^2+dy^2+dz^2),$$
the total system of dynamic equations \eqref{Eq__3} and \eqref{Eq__4} for the scale factor \emph{a(t)}   and the scalar potential $\Phi(t)$  can be written in the form
\begin{equation} \label{Eq__7}
\ddot{\Phi}=-3H\dot{\Phi}-em^2\Phi+\alpha\Phi^3
\end{equation}
\begin{equation} \label{Eq__8}
3H^2-\mathcal{E}=0
\end{equation}
\begin{equation} \label{Eq__9}
\dot{H}=-3H^2+\frac{em^2\Phi^2}{2}-\frac{\alpha\Phi^4}{4}+\lambda
\end{equation}
where
\begin{equation} \label{Eq__10}
H(t)=\frac{\dot{a}}{a}
\end{equation}
is the Hubble constant,
\begin{equation} \label{Eq__11}
\mathcal{E}=\frac{\dot{\Phi}^2}{2}+\frac{em^2\Phi^2}{2}-\frac{\alpha\Phi^4}{4}+\lambda
\end{equation}
is the \emph{effective energy} of the system which, according to \eqref{Eq__8}, is a non-negative value:
\begin{equation} \label{Eq__12}
\mathcal{E}\geqslant 0.
\end{equation}
Substitution of the Einstein equation \eqref{Eq__8} into the right-hand side of  \eqref{Eq__9} gives the relationship
\begin{equation} \label{Eq__13}
\dot{H}=-\frac{1}{2}\dot{\Phi}^2 (\leqslant 0),
\end{equation}
according to which the Hubble  constant in the model can only decrease with time and $H=Const\Leftrightarrow \dot{\Phi}=0$. Introducing also the \emph{invariant cosmological acceleration}
\begin{equation} \label{Eq__14}
\Omega\equiv \frac{a\ddot{a}}{\dot{a}^2}\equiv 1+\frac{\dot{H}}{H^2},
\end{equation}
we represent\eqref{Eq__13} in the equivalent form:
\begin{equation} \label{Eq__15}
\Omega=1-\frac{1}{2}\frac{\dot{H}}{H^2}(\leqslant 1).
\end{equation}
We also introduce a useful relationship for the quadratic invariant of the Friedman space curvature %
\begin{equation} \label{Eq__16}
\sigma\equiv \sqrt{R_{ijkl}R^{ijkl}}=H^2\sqrt{6(1+\Omega^2)}\geqslant 0.
\end{equation}
Introducing then the dimensionless time variable $\tau=mt$, $\phi'=d\phi/d\tau$ , the dimensionless fundamental constants, and the dimensionless functions
\
\begin{eqnarray}\label{Eq__17}
\alpha_m=\frac{\alpha}{m^2}; \quad \lambda_m=\frac{\lambda}{m^2};
\quad h=\frac{a'}{a}=\frac{H}{m};
\end{eqnarray}
we rewrite system of basic equations \eqref{Eq__7} - \eqref{Eq__9}
in the normal form for variables $\Phi\equiv x_1, Z\equiv x_2$, and $h\equiv x_3$ :
\begin{eqnarray} \label{Eq__18}
\Phi'=Z & (\equiv f_1),\\
\label{Eq__19}
Z'=-3hZ-e\Phi+\alpha_m\Phi^3 & (\equiv f_2),\\
\label{Eq__20}
h'=-3h^2+\frac{e\Phi^2}{2}-\frac{\alpha_m\Phi^4}{4}+\lambda_m &(\equiv f_3)
\end{eqnarray}

In this case, expressions \eqref{Eq__14} and \eqref{Eq__15} for invariant cosmological accelerations remain invariant after the replacement of the derivatives $H\rightarrow h$ and $d/dt\rightarrow d/d\tau$, and expression \eqref{Eq__11} for the effective energy is transformed as follows:
\begin{eqnarray} \label{Eq__21}
\mathcal{E}=m^2\mathcal{E}_m, & \mathcal{E}_m(\Phi,Z)=\displaystyle\frac{Z^2}{2}+\frac{e\Phi^2}{2}-\frac{\alpha_m\Phi^4}{4}+\lambda_m\geqslant 0.
\end{eqnarray}
Let us write down in these designations integral condition \eqref{Eq__8}:
\begin{equation} \label{Eq__22}
3h^2-\frac{Z^2}{2}-\frac{e\Phi^2}{2}+\frac{\alpha_m\Phi^4}{4}-\lambda_m=0
\end{equation}
Note that all variables and constants in our designations are dimensionless. So, the dynamic system under study has three degrees of freedom, and its state is unambiguously defined by the coordinates of the point $M(\Phi,Z,h)\equiv M(x_1,x_2,x_3)$  in the 3-dimensional phase space $\mathbb{R}^3$ . Moreover, regions of the phase space $\Omega\subset \mathbb{R}^3$  in which energy condition \eqref{Eq__12} is violated, that is,
\begin{eqnarray} \label{Eq__23}
\Omega\subset \mathbb{R}^3: & \mathcal{E}_m(\Phi,Z)<0
\end{eqnarray}
are inaccessible to the dynamic system; as a result, the phase space of the dynamic system can appear multi connected. Note that inaccessible regions \eqref{Eq__23}, if they exist, are cylindrical with the axes parallel to \emph{Oh} . Furthermore, Einstein's equation \eqref{Eq__22} being \emph{the first integral of the dynamic system} - the integral with the zero total energy - describes at the same time the fourth order algebraic surface in the phase space $\mathbb{R}^3$   of the dynamic system $\sum\subset\mathbb{R}^3$  in which all phase trajectories of the dynamic system lie. Following \cite{Ign5}, we call the surface $\sum$ below the \emph{Einstein-Higgs} hypersurface.
\subsection{Singular points of the dynamic system}
The singular points of the dynamic system are defined by equality to zero of the right hand sides of the normal system of equations. Thus, the coordinates of the singular points of dynamic system \eqref{Eq__18}-\eqref{Eq__20} are defined by the equations
\begin{eqnarray} \label{Eq__24}
Z=0, & -3hZ-e\Phi+\alpha_m\Phi^3=0,
\end{eqnarray}
$$-3h^2+\frac{e\Phi^2}{2}-\frac{\alpha_m\Phi^4}{4}+\lambda_m=0$$
From here we generally obtain six singular points of the system - two symmetric points with zero potential and its derivative \cite{Ign5}:
\begin{eqnarray} \label{Eq__25}
M_{\pm}\bigg(0,0,\pm \sqrt{\frac{\lambda_m}{3}}\bigg) & (\textrm{for}\textrm{ }\lambda_m\geqslant 0, \forall \alpha)
\end{eqnarray}
and four symmetric points $M_{ab}$ located at vertices of the rectangle in the plane $P_2(\Phi,Z):Z=0:$
\begin{eqnarray} \label{Eq__26}
M_{ab}\bigg(\pm\frac{1}{\sqrt{e\alpha_m}},0,\pm \sqrt{\frac{\lambda_\alpha}{3}}\bigg) & \textrm{for}\textrm{ }e\alpha>0,\textrm{ }\lambda_\alpha\equiv \lambda_m+\displaystyle\frac{1}{4\alpha_m}\geqslant 0.
\end{eqnarray}
Thus, all five cases \cite{Ign5} are possible:
\begin{enumerate}
  \item for $\{e\alpha<0,\lambda<0\}$, no singular points exist,
  \item for $\{e\alpha>0,\lambda<-1/4e\alpha_m\}$, no singular points exist,
  \item for $\{e\alpha\langle0,\lambda\rangle0\}$, only two singular points $M_\pm$ exist,
  \item for $\{e\alpha>0,-1/4\alpha_m<\lambda<0\}$, four singular points $M_{11},M_{12},M_{21}$ and $M_{22}$ exist,
  \item for $\{e\alpha>0,\lambda>0\}$, all six singular l points exist.
\end{enumerate}
Calculating effective energy \eqref{Eq__21} at the singular points, we obtain
\begin{eqnarray} \label{Eq__27}
\mathcal{E}_m(M_\pm)=\lambda_m, & \mathcal{E}_m(M_{ab})=\lambda_\alpha
\end{eqnarray}
So, from conditions \eqref{Eq__25} and \eqref{Eq__26} it follows that \emph{all singular  points of the dynamic system, if they exist, are accessible}. The eigenvalues of the dynamic system matrix at the singular points are \cite{Ign5}
\begin{eqnarray}
M_\pm: & k_1=\mp2\sqrt{3\lambda_m}, & k_{2,3}=\mp\frac{1}{2}\sqrt{3\lambda_m}\pm\frac{1}{2}\sqrt{3\lambda_m-4e},\nonumber\\
M_{ab}: & k_1=\mp2\sqrt{3\lambda_\lambda}, & k_{2,3}=\mp\frac{1}{2}\sqrt{3\lambda_\lambda}\pm\frac{1}{2}\sqrt{3\lambda_\alpha-8e}\nonumber
\end{eqnarray}
For pairs of the symmetric points $(M_{11},M_{21})$ and $(M_{12},M_{22})$ , the eigenvalues of the dynamic system matrix coincide. The signs before the first and second radicals in these formulas take values independently from each other: the signs before the first radicals correspond to different pairs of points, and the signs before the second radicals - to different eigenvalues. In this case, it is important to remember necessary conditions \eqref{Eq__25} and \eqref{Eq__26} of existence of the singular points.

Thus, the singular points $M_\pm$  for $3\lambda_m-4e>0$  are unstable (saddle) points, and for $3\lambda_m-4e<0$, they are stable points (attraction centers). At these points, the Hubble constant $h=\pm\sqrt{\lambda_m/3}$ for $\lambda>0$, and the scalar field is absent. According to formulas \eqref{Eq__21} and \eqref{Eq__23}, the invariant cosmological acceleration and the invariant curvature are
\begin{eqnarray} \label{Eq__28}
\Omega(M_\pm)=1,& \sigma(M_\pm)=\displaystyle\frac{2}{\sqrt{3}}\lambda
\end{eqnarray}
As a result, inflationary solutions with positive or negative Hubble constant $a\sim e^{\pm\sqrt{\lambda/3t}}$ correspond to the singular points $M_\pm$. For  $e=+1$, all eigenvalues of the dynamic system matrix at its singular points  $M_{ab}$ are real and have different signs. Thus, for  $e=+1$, all singular points are saddle points. For $e=-1$  and $\lambda_\alpha<8/3$, the values of the dynamic matrix at its singular points $M_{ab}$ become complex conjugated, and among them there are values corresponding to the attraction.
\subsection{Asymptotic phase trajectories}

Dynamic system of equations \eqref{Eq__18}-\eqref{Eq__20} is unambiguously set by the \emph{dynamic system matrix}
\begin{equation} \label{MatrixSystem}
\textbf{A}=
\begin{Vmatrix}
\displaystyle\frac{\partial f_i}{\partial f_k}
\end{Vmatrix},\nonumber
\end{equation}
and its phase trajectories in the vicinities of the singular points  $ M^0(x^0_i)$ are determined by asymptotic formulas (for example, see \cite{Bogoyav})
\begin{equation} \label{Eq__29}
r(t)=r^0+\textrm{Re}\sum^{n}_{k=1}C_k\xi^ke^{\lambda_kt}
\end{equation}
where $\xi^k$ are eigenvectors corresponding to the eigenvalues $\lambda_k$ of the matrix $\textbf{A}$:
\begin{equation} \label{Eq__30}
(\textbf{A}-\lambda\textbf{\textbf{E}})\xi=0.
\end{equation}	
To determine the constants $C_i$, the initial conditions $r(t_0)\equiv r_0$  must be assigned at a point $M_0$ in close vicinity of  the singular point $M^0$. Examples of such trajectories for the fundamental parameters $\alpha_m=5$, $e=+1$ and $\lambda=0.1$ are shown in Figs. 1 and 2. The singular points are designated by filled circles, the initial points are shown by asterisks, the beginning of the trajectories is indicated by the circle/sphere, and the end is indicated by the square/cube.
\section{Einstein-Higgs hypersurfaces}
\subsection{Surface topology and form of the cross sections}
According to formula \eqref{Eq__13}, $h\leqslant 0$; therefore, the Hubble constant cannot increase with time. The possibility of transition from the expansion ($h>0$) to the compression stage ($h<0$ ) is defined by the Einstein-Higgs hypersurface topology. We now rewrite equation \eqref{Eq__22} for the Einstein-Higgs hypersurface in the equivalent form
\begin{eqnarray} \label{Eq__31}
\sum:3h^2-\frac{Z^2}{2}+\frac{\alpha_m}{4}\bigg(\Phi^2-\frac{e}{\alpha_m}\bigg)^2=\lambda_\alpha &
\bigg(\lambda_\alpha=\lambda+\displaystyle\frac{1}{4\alpha_m}\bigg)
\end{eqnarray}
Note that according to \eqref{Eq__5}, $\lambda_\alpha=\lambda_0/m^2$. After the replacement
\begin{equation} \label{Eq__32}
\Phi^2-\frac{e}{\alpha_m}=y,
\end{equation}	
we reduce the equation for the Einstein-Higgs hypersurface to the equation of the second-order \emph{central} surface in the three-dimensional space:
$$3h^2-\frac{Z^2}{2}+\frac{\alpha_my^2}{4}=\lambda_\alpha$$
For $\alpha_m>0$, its main axis is the $OZ$ axis, and for $\alpha<0$, it is the $Oh$  axis. At  $\lambda_0=0$, the Einstein-Higgs hypersurface is a cone in the variables $\{y,Z,h\}$. Assuming that $\lambda_0\neq 0$, we reduce the equation for the Einstein-Higgs hypersurface to the canonical form
\begin{equation} \label{Eq__33}
\frac{3h^2}{\lambda_\alpha}-\frac{Z^2}{2\lambda_\alpha}+\frac{\alpha_m^2y^2}{4\beta}=1,
\end{equation}	
where  $\beta=\alpha_m\lambda_\alpha\equiv 1/4+\alpha_m\lambda$. Thus, the type of the Einstein-Higss surface described by \eqref{Eq__33} in the variables  $\{y,Z,h\}$ is defined by the sign of  $\beta$: for  $\beta>0$, it is a one-sheeted hyperboloid, at  $\beta=0$, it is a cone, and for $\beta<0$, it is a two-sheeted hyperboloid. In this case, we should bear in mind that transformation \eqref{Eq__32} is not bijective: for the inverse transformation, we have two roots  $\Phi=\pm\sqrt{y+e/\alpha_m}$, and the second-order surface is deformed. The greatest deformations arise in the region of singular points $|\Phi|\leqslant 1$, $|\Phi|\sim 1/\sqrt{e\alpha_m}$. In this region, the cross section of the Einstein-Higgs surface can have quite diverse forms largely defining the possibilities of transition of the cosmological model from the stage of expansion to the stage of compression. In particular, the cross section $\textbf{S}:Z=0$ (that is, in the $\{\Phi,h\}$  plane) in which the singular points lie is determined by the equation
$$\textbf{S}:\frac{3h^2}{\lambda_\alpha}+\frac{\alpha_\alpha}{4\lambda\alpha}\bigg(\Phi^2-\frac{e}{\alpha}\bigg)^2=1$$
according to which the following forms of the section are possible:  $\beta<0\Rightarrow \textbf{S}=\varnothing$,  $\beta<0\Rightarrow \textbf{S}=M_1\cup M_2$,  $0<\beta<1/4\Rightarrow \textbf{S}=\textbf{E}_1\cup \textbf{E}_2$, and  $\beta>1/4\Rightarrow \textbf{S}=\textbf{E}$, or $\infty$ where $M_a$ are the singular points,  $\textbf{E}_a$ are the deformed ellipses, and $\infty$  denotes the infinity.
\TwoFig{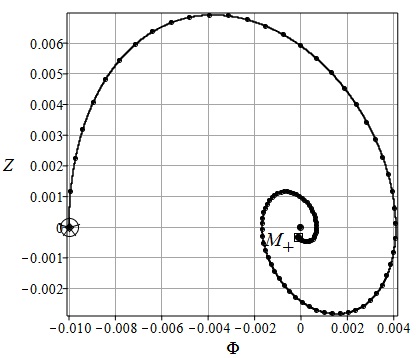}{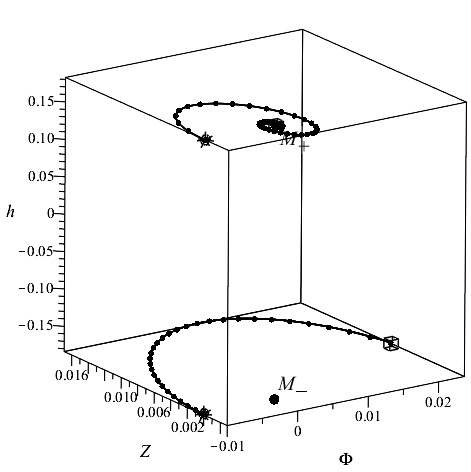}{\label{Fig1}
Phase trajectory of dynamic system \eqref{Eq__18}-\eqref{Eq__20}
(the solid curve) and the asymptotic trajectory (points) on the  $\{\Phi Z\}$ plane}
{\label{Fig2} Phase trajectories of dynamic system \eqref{Eq__18}-\eqref{Eq__20}
(the solid curve) and asymptotic trajectory (points) in the 3-dimensional $\{\Phi Z\}$  space.}

\subsection{Types of the Einstein-Higgs hypersurface and its cross sections}
Discussing the types of the Einstein-Higgs hypersurfaces according to the classification of the central surfaces (Figs. \ref{Fig3}-\ref{Fig10}), by default we mean the adjective \emph{deformed}.
\TwoFigName{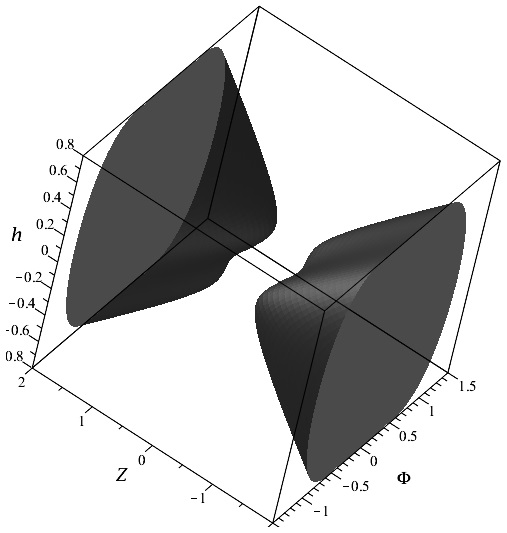}{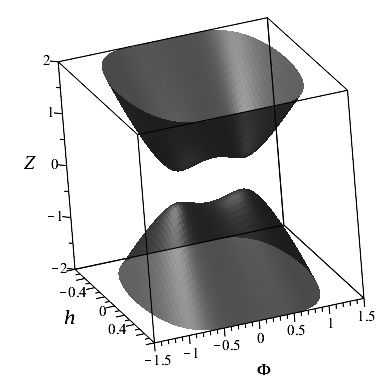}{\label{Fig3}
Two-sheeted hyperboloid with $\alpha_m=5, e=+1$  and $\lambda=-0.1\Rightarrow \beta=-1/4$. The principal axis is $OZ$.}
\ThreeFigName {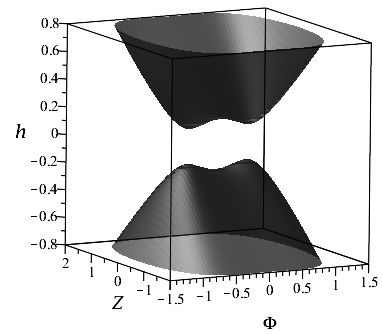}{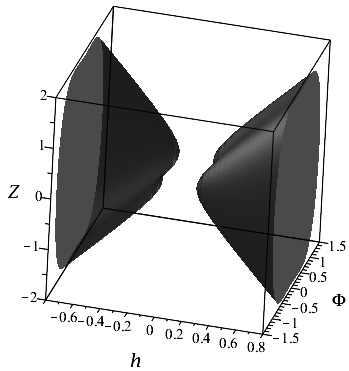}{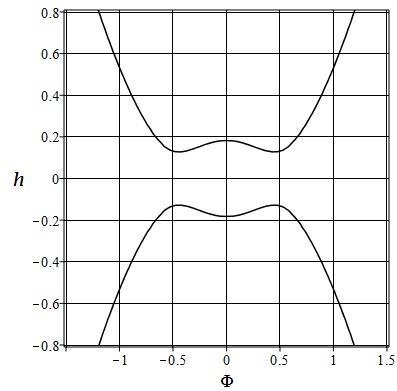}{\label{Fig4}
Two-sheeted hyperboloid with $\alpha_m=-5, e=-1$  and $\lambda=-0.1\Rightarrow \beta=+1/4$. The principal axis is  $Oh$.}

From these examples it can be seen that the surface topology in Fig. \ref{Fig3} admits transitions of the cosmological models from the expansion to the compression stage
 $H_+\rightarrow H_-$, and the surface topology in Fig. \ref{Fig4} does not admit it.
\ThreeFigName {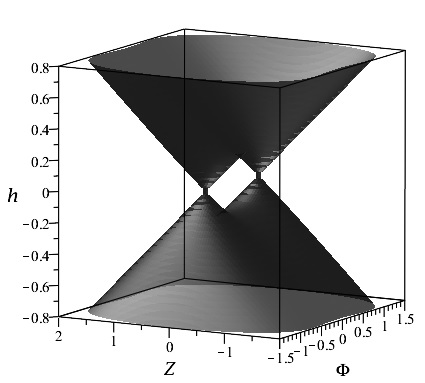}{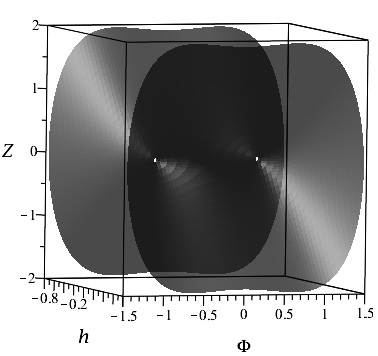}{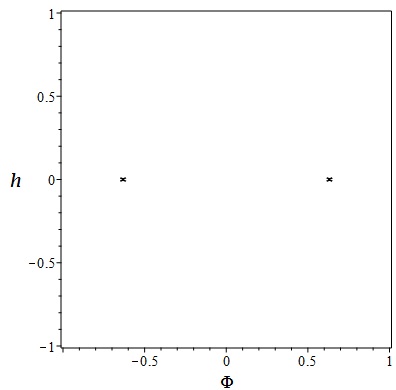}{\label{Fig5}
Cone with  $\alpha_m=-2.5, e=+1$  and $\lambda=-0.1\Rightarrow \beta=0$.
The principal axis is $Oh$. The cross section points  $h=0$ are also indicated.}
\TwoFigName{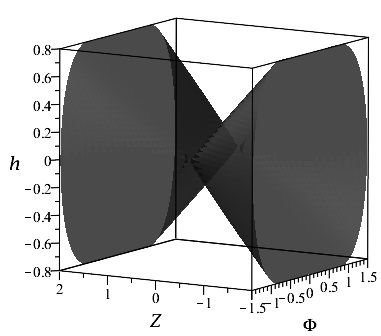}{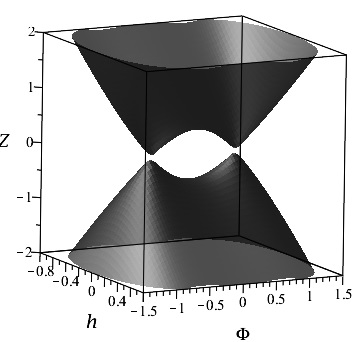}{\label{Fig6}
Cone with $\alpha_m=2.5, e=+1$  and $\lambda=-0.1\Rightarrow \beta=0$. The principal axis is  $OZ$.}
\ThreeFigName {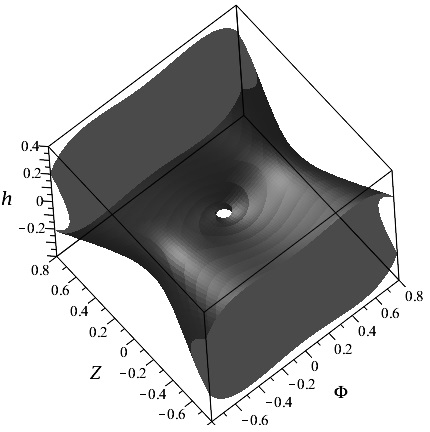}{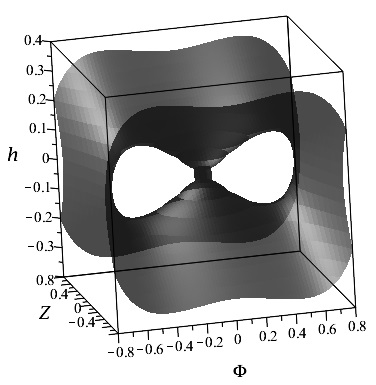}{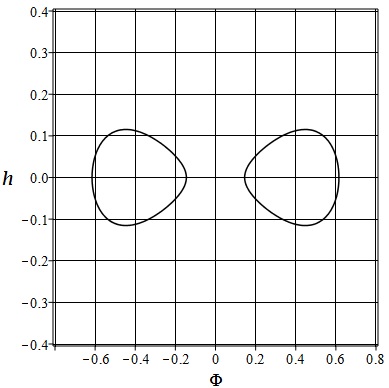}{\label{Fig7}
One-sheeted hyperboloid with  $\alpha_m=5, e=+1$  and $\lambda=-0.01\Rightarrow \beta=0.2$. The principal axis is $OZ$.}
\ThreeFigName {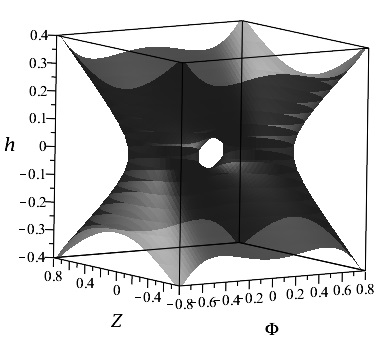}{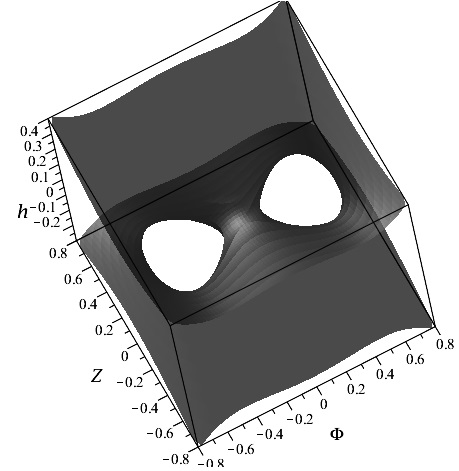}{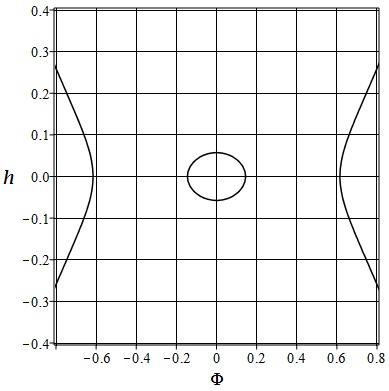}{\label{Fig8}
One-sheeted hyperboloid with  $\alpha_m=-5, e=-1$  and $\lambda=-0.01\Rightarrow \beta=0.2$. The principal axis is  $OZ$.}
\ThreeFigName {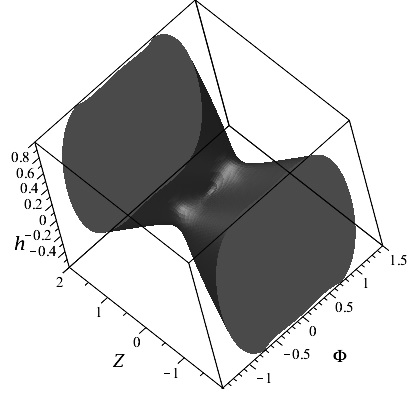}{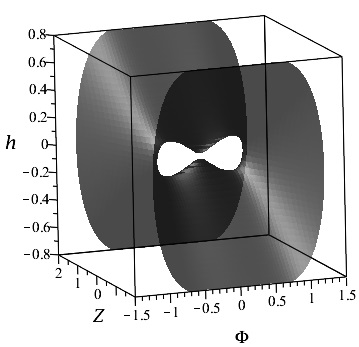}{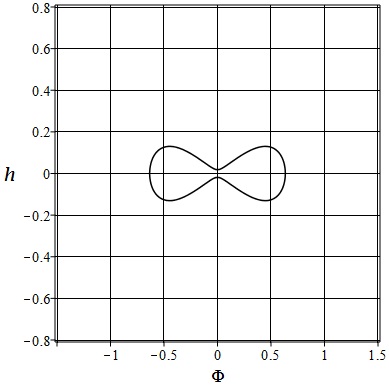}{\label{Fig9}
One-sheeted hyperboloid with  $\alpha_m=2, e=+1$  and $\lambda=0.001\Rightarrow \beta=0.252>1/4$. The principal axis is  $OZ$.}
\ThreeFigName {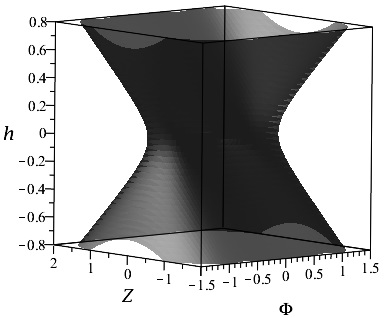}{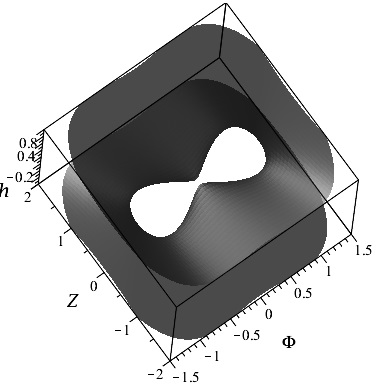}{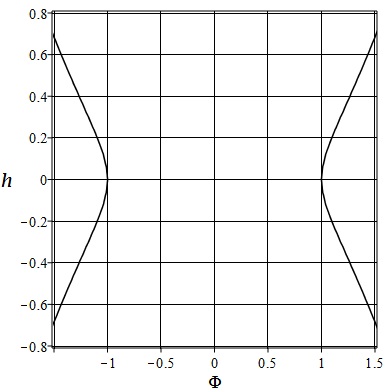}{\label{Fig10}
One-sheeted hyperboloid. The principal axis is $Oh$.
In Part II of the paper, we will discuss results of numerical simulation of the cosmological model presented above.}
The Author is thankful to participants of Kazan University's seminar of the department of relativity theory and gravitation for helpful discussion of this work\footnote{This paper has been supported by the Kazan Federal University Strategic Academic Leadership Program..}.
%

\vfill


\begin{thebibliography}{90}
\vspace{3mm}
\bibitem{IgnSTFI}
Yu. G. Ignat'ev, Space, Time and Fundamental Interactions, No. 2, 36 (2017).
\bibitem{IgnRPhJ}
Yu. G. Ignatyev, Russ. Phys. J, 59, No. 12, 2074-2079 (2016).
\bibitem{Ign1}
Yu.G. Ignat'ev and I. A. Kokh, Gravit. Cosmol., 25, 24 (2019).
\bibitem{Ign2}
Yu. I. Ignat'ev and I. A. Kokh, Gravit. Cosmol., 25, 37 (2019).
\bibitem{Ign3}
Yurii Ignat'ev, Alexander Agathonov, and Irina Kokh, arXiv:1810.09873 [gr-qs] (2018).
\bibitem{IgnSam1}
Yu. G. Ignat'ev and A. R. Samigullina, Russ. Phys. J., 61, No. 4, 643-647 (2018).
\bibitem{Ign4}
Yu. G. Ignat'ev and I. A. Kokh, Russ. Phys. J., 61, No. 9, 1590-1596 (2018).
\bibitem{IgnSam2}
Yu. G. Ignatyev and A. R. Samigullina, Russ. Phys. J., 62, No. 4, 618-626 (2019).
\bibitem{IgnSam3}
Yu. G. Ignat'ev and A. R. Samigullina, Russ. Phys. J., 63, No. 1, 23-33 (2020).
\bibitem{Ign5}
Yu. G. Ignat'ev and D. Yu. Ignatyev, Gravit. Cosmol., 26, 29 (2020).
\bibitem{Bogoyav}
O. I. Bogoyavlenskii, Methods of Qualitative Theory of Dynamic Systems in Astrophysics and Gas Dynamics [in Russian], Nauka, Moscow (1980).
%
\end{thebibliography}
\end{document}